\documentclass{article}

\usepackage[utf8]{inputenc}
\usepackage[T1]{fontenc}
\usepackage{hyperref}
\usepackage{url}
\usepackage{booktabs}
\usepackage{amsfonts}
\usepackage{nicefrac}
\usepackage{microtype}
\usepackage{graphicx}
\usepackage{color}
\usepackage{orcidlink}

\title{Communication Skills in Software Engineering: A Multivocal Review}

\author{
  Dannilo Rabelo\orcidlink{0000-0002-4047-2657} \\
  Universidade Federal do Pará \\
  \texttt{dannilo.cabral@gmail.com} \\
  \and
  Deisy Peres\orcidlink{0009-0007-6552-8593} \\
  Universidade Federal de Pernambuco \\
  \texttt{deisyperes@gmail.com} \\
  \and
  Emmanuel Dias\orcidlink{0000-0002-0921-8992} \\
  Universidade Federal do Pará \\
  \texttt{emmanueldias@defensoria.apdef.br} \\
  \and
  Thayssa Rocha\orcidlink{0000-0001-8026-4003} \\
  Universidade Federal do Pará \\
  \texttt{thayssa.rocha@icen.ufpa.br} \\
  \and
  Enne Rebeca de Freitas\orcidlink{0009-0006-2244-0754} \\
  Universidade Federal do Pará \\
  \texttt{enne.freitas@icsa.ufpa.br} \\
  \and
  Kiev Gama\orcidlink{0000-0003-1508-6196} \\
  Universidade Federal de Pernambuco \\
  \texttt{kiev.gama@gmail.com} \\
  \and
  Gustavo Pinto\orcidlink{0000-0003-4900-6369} \\
  Universidade Federal do Pará \\
  \texttt{gpinto@ufpa.br} \\
}

\begin{document}

\maketitle

\begin{abstract}
Communication skills are increasingly recognized as essential in Software Engineering, yet discussions about them remain fragmented across academic and gray literature. This fragmentation is problematic because it limits a broader understanding of how communication is valued, taught, and applied in both educational and professional settings. Through a multivocal literature review, we found strong convergence between academic and gray sources in treating communication as a core competency, while also identifying differences in emphasis, with academia focusing on conceptualization and empirical evidence and gray literature stressing practical consequences and emerging industry practices.
\end{abstract}

\section{Introduction}

Traditionally, Software Engineering (SE) emphasizes the mastery of technical skills (e.g., programming, algorithms, system design). However, over the years, a new trend has emerged, demonstrating that the success of software projects depends not only on technical abilities but also on the human factor and interpersonal skills \cite{sommerville2011software}, among which communication holds a prominent position\cite{borges2024skills}. Communication in this context refers to a range of competencies that encompass the ability to clearly and effectively explain, transmit, and receive messages where requirements are fully elucidated~\cite{schwaber2011scrum}.
In modern and multidisciplinary development teams, the effective exchange of information is vital for task coordination, conflict resolution, and ensuring that everyone works with a shared understanding of the project objectives\cite{adham2023conflict}.

This primacy of communication has been widely debated in both academia and the industry. On the one hand, researchers often fail to consider the daily challenges faced in development teams; on the other, industry professionals are unaware of research findings that could guide them in improving processes and developing competencies.
Since in SE the human competencies and heavily context-dependent, it is essential that industry practice provides concrete examples, but also that academia contributes structured models for their development and evaluation.
Given this scenario, the purpose of this work is to bridge these two sources of knowledge, seeking to answer the following research questions:
\begin{itemize}
    \item \textbf{RQ1:} How are communication skills in Software Engineering discussed and valued across different sources (academic and gray literature)?
    \item \textbf{RQ2:} Which divergences and/or convergences exist between the perspectives of the academic literature and the gray literature on the topic?
\end{itemize}

To address \textbf{RQ1}, we examined representative academic studies and industry reports that discuss the importance of communication in the SE context. This allowed us to identify the main arguments, evidence, and focuses of each domain. Subsequently, for \textbf{RQ2}, we compared the findings from both sources to highlight convergences and divergences in how each community perceives and values communication.

\section{Communication as a Core Competency in Software Engineering}

Recent studies in SE treats communication as a core professional competency, often alongside technical expertise. Studies based on job advertisements\cite{matturro2013soft, rabelo2022, galster2022, florea2018software}, systematic reviews\cite{matturro2019systematic}, and employer perceptions show that communication, teamwork, and collaboration are perceived as demanded skills in the software industry. This evidence suggests that communication is not peripheral, but a central condition for effective professional performance and project success.

Academic work also highlights the educational dimension of this problem. Researchers argue that Software Engineering courses have historically underemphasized communication training, despite market expectations for oral, written, and collaborative abilities \cite{liu2005teaching,scaffidi2018survey}. More recent studies propose breaking communication down into sub-skills such as technical writing, active listening, intercultural competence, and negotiation, which can then be taught and assessed more systematically\cite{kapur2020types}.

Gray literature reinforces the same core message, but from a more practical and immediate perspective. Blogs, company reports, and professional platforms frame communication as essential for collaboration, leadership, promotion, and day-to-day engineering work, including documentation, stakeholder alignment, and feedback exchange~\cite{medium2023}. At the same time, gray literature appears to anticipate emerging practices such as asynchronous communication, technical storytelling, and living documentation, which are still less visible in formal academic research~\cite{garousi2022introduction,zarb2014evaluating,choi2009pair}.

\section{Research Methodology}

To answer the proposed research questions, we conducted a Multivocal Literature Review (MLR)~\cite{GAROUSI2019}, integrating evidence from both academic literature and gray literature sources.
The research process was divided into three stages: (1) Definition of research question; (2) Search and select studies; (3) Extracting and synthesizing data.

The first step, called \textbf{definition of research question}, initially involves delimited narrowing the scope of the investigation to focus specifically on communication skills in SE. We improved RQ1 to encompass not only whether communication is important (which the literature already indicated), but also how it is discussed and valued, that is, which specific aspects of communication are mentioned, in which contexts (teamwork, customer relationships, etc.), and what evidence is available.

For RQ2, we defined "databases" as synonymous with knowledge bases or sets of sources, in this case, including formal academic literature versus industry grey literature.

For the second stage \textbf{Search and select studies}, we conducted searches in four scientific databases between January 2024 and March 2024.
(IEEE Xplore, ACM Digital Library, Scopus, Google Scholar) using keywords such as ``software engineering communication skills'', ``soft skills software developers'', ``software engineering teamwork communication''. Over the results we applied filters for studies published mainly in the last decade (2013 to 2023), but without excluding older seminal references. We selected relevant research articles, including: empirical studies on skills demand (analysis of job advertisements~\cite{matturro2013soft, florea2018software}), systematic reviews on soft skills~\cite{matturro2019systematic}, and specific research on communication (e.g., studies on collaboration in distributed teams, communication teaching).

In the third stage, called \textbf{extracting and synthesizing data}
For each selected source (both academic and gray), we performed an analytical reading, extracting the central points related to communication skills.
In the formal academic literature, we reviewed 15 key academic articles, ensuring geographical diversification (studies in South America, North America, Europe, and Oceania) and focus (industry, education, and agile methods). We prioritized recent and/or highly referenced content within the community.

To embrace the gray literature we conducted searches on Google and in repositories of technical/popular blogs using the terms: e.g. ``importance of communication software developer'', ``software engineer communication soft skill'', ``how software engineers learn communication''. We identified influential blog posts, articles on IT career sites, discussions on the Stack Overflow~\footnote{A question and answer website for professionals and enthusiasts in the field of computer programming. https://stackoverflow.com/questions} Blog and Medium, and reports from organizations (e.g., research by the Association of American Colleges \& Universities cited via Inc.com).
10 items of gray literature were included, among them: personal blogs of senior developers (e.g., Chintan Khetiya, 2021), corporate articles (Stack Overflow, 2022; WeCP, 2025), and opinion articles from specialized websites. We ensured that the chosen sources directly addressed the relevance of communication or offered practical examples related to SE.

For each selected source (academic and non-academic), extracting the central points related to communication skills.

In academic literature, the focus has been on the results and conclusions of studies, for example, whether the article listed "communication" as one of the most important skills, and what the evidence was (percentage of job openings, interviews, etc.), and in what context (developer, tester, manager roles). In the gray literature, we extracted key statements (e.g., ``communication is the most important skill'') and the justifications provided (why the blog author believes this, what examples were given). This information was organized into a categorized structure, where we identified dimensions such as the relative importance of communication, valued types of communication (oral/written, internal/external), effects of poor communication, tips for improvement, etc. This categorization helped group similar findings and also highlight differences in emphasis between sources. We used a spreadsheet to tabulate the findings, with separate columns for source, type (academic/gray), and extracted points. All the considered papers and articles from the grey literature, as well as the categorization construction spreadsheet, can be found at the supplementary information repository \footnote{https://figshare.com/s/13ffc6100bb9539d22e1}.

Data selection was performed by the first author and validated by the last author. The data extraction from the selected sources and subsequent coding and categorization were performed manually by the first and last author.

As a methodological limitation, it is worth noting that the selection of the gray literature, while careful, does not guarantee complete coverage, as the universe of blogs and posts is virtually infinite and dynamic. We sought to mitigate this by choosing widely read sources that resonate with frequent topics in communities (measured by shares, comments, etc.). In the academic literature, our search was limited on English written papers. Nevertheless, we believe the collected sample captured the main perspectives and answer \textbf{RQ1} and \textbf{RQ2}.

\section{Results}

The analysis of the results reveals that both the academic and gray literatures converge in recognizing communication as a central competency, but with distinct focuses. Academia concentrates on teaching methodologies, quantitative measurement, and curricular integration, empirically demonstrating how communication affects learning and technical performance. The gray literature, in contrast, emphasizes immediate practical consequences, such as the risk of dismissal, lack of promotion, or rework, in addition to bringing up emerging industry practices like storytelling and continuous feedback.

\subsection{RQ1: How are communication skills in SE discussed and valued?}

The synthesis of the studied sources reveals a strong consensus on the crucial importance of communication skills in Software Engineering. Both academic researchers and industry professionals describe communication as an indispensable component of the software engineer's profile, although they use slightly different voices and focuses. Below, we summarize the main points observed:

\paragraph{\textbf{Communication as a Top Soft Skill:}} Several formal studies place communication at the top or among the first places in the lists of soft skills demanded by the software market. Florea \& Stray\cite{florea2018software}, when analyzing job openings for testers globally, found that communication was the most cited skill category by employers, surpassing others like problem-solving and teamwork. Matturro\cite{matturro2013soft} identified effective oral communication as one of the five most relevant skills in IT job postings in Uruguay. In the gray literature, this prioritization appears assertively: blogs frequently proclaim communication as ``the most important skill'' or ``the \#1 characteristic'' for professional success, even above technical talents.

\paragraph{\textbf{Justifications for Importance:}} The sources converge in explaining why communication is so vital. The most cited reason is that software development is inherently a collaborative effort. Projects are rarely conducted by isolated individuals; on the contrary, they require the coordination of multidisciplinary teams, interaction with clients and users, and the integration of different modules and ideas. Thus, communicating well becomes a condition for aligning expectations, avoiding rework, and ensuring cohesion in the final product. Academic studies emphasize that teams with better internal communication tend to have fewer coordination failures and higher productivity, especially in agile methodologies where information flows quickly in daily stand-ups and reviews. On the industry side, it is frequently cited that communication failures generate conflicts, delays, and loss of quality—"misunderstandings can derail projects," while professionals with good communication can facilitate discussions and group problem-solving.

\paragraph{\textbf{Valued Different Facets of Communication:}} Communication in SE is not treated as a monolith; several specific facets are discussed. Oral communication (spoken) is vital in meetings, presentations, and daily interpersonal interactions. Written communication is highlighted in relation to software documentation, writing clear emails, reports, and using collaborative tools (wikis, tickets) comprehensibly. For example, industry sources point out that an engineer who "excels at written communication can produce documentation that is not only informative but easy to understand even for non-engineers."
Active listening is another sub-skill emphasized, especially in the context of teamwork: knowing how to listen to colleagues and clients to understand needs and concerns. Some materials, primarily academic ones, mention foreign language communication (English) as crucial in globalized development environments where teams and clients are international. Finally, the ability to give and receive constructive feedback appears as a highly cited communication attribute, as in modern practices (e.g., code review, agile retrospectives), developers are expected to professionally critique each other's work and incorporate criticism without personal friction.

\paragraph{\textbf{Consequences of Communication Failure:}}
Development teams suffer from loss of information, which can result in incorrectly implemented functionalities, bugs that take time to be identified due to lack of reporting, or even security failures when requirements are not clearly transmitted. Industry blogs mention ``telephone game'' scenarios where a failure to clarify doubts led to months of work in the wrong direction. From a personal standpoint, developers with communication difficulties can become isolated within the team, failing to share knowledge (which harms both them—as they miss out on recognition—and the team—which loses opportunities to reuse solutions). This isolation can also generate friction: e.g., a developer who cannot explain their design might receive criticism and take it personally, creating a tense environment. Several blog authors explicitly recommend: "if you want to avoid unnecessary conflict and stress, improve your communication," pointing out that many technical conflicts are, in fact, communication problems.

In summary, the answer to RQ1 is that communication skills are unanimously considered of enormous importance in Software Engineering, receiving attention in both scientific studies and practical discussions. They are valued for their direct contribution to effective collaboration, clarity in defining and meeting requirements, and a good organizational climate.

\subsection{RQ2: Convergences and Divergences Between Academic and Gray Literature}

Upon confronting the findings of the two knowledge bases examined, we identified a predominantly convergent landscape: both domains explicitly recognize the critical value of communication skills in Software Engineering. However, subtle divergences also emerge in how the topic is approached and elaborated. We can summarize the answer to RQ2 in the following points:

\subsubsection{\textbf{Convergences (Points of Agreement):}}

\paragraph{\textbf{Communication Drives Collaboration and Effectiveness:}} One point of agreement is the view that communication enables other essential practices, such as effective teamwork, collective problem-solving, and aligning expectations with stakeholders. Academic sources quantify this by showing correlations or the concurrent frequency of mention of ``communication'' and ``teamwork'' in job requirements. Industry sources report this connection in the form of stories and maxims ("good communication = cohesive team"). In short, there is consensus that communication is the foundation for collaboration (and the two words often appear together in discussions, reinforcing this conceptual convergence).

\paragraph{\textbf{Need to Develop Communication Skills:}} Both researchers and practitioners advocate for the necessity of improving software engineers' communication skills. There is a mutual recognition that not all professionals naturally arrive with this competence well-formed, and that deliberate efforts are required. Academia proposes educational interventions (communication courses, use of active methodologies focused on soft skills, etc.), while the industry suggests in-service training, mentorship, and behavioral assessments. Although the strategies vary, both agree on the diagnosis: traditional education has not placed sufficient emphasis, so actions are welcome to bridge this gap.

\subsubsection{\textbf{Divergences (Differences and Distinct Emphases):}}

\paragraph{\textbf{Empirical vs. Tacit Approach:}} A marked difference lies in the style of evidence. The academic literature treats the importance of communication empirically and analytically: it presents survey data, job vacancy percentages, results of educational experiments, etc. In contrast, the gray literature often relies on personal experiences, anecdotal cases, and general advice. For example, to state that communication is crucial, an academic article might say, ``X\% of job advertisements require communication, the second most cited skill,'' while a blog will say, ``if you don't communicate well, you won't be promoted; look at cases of people fired for it.'' Both convey the same message, but the form of argumentation differs: the former appeals to the authority of numbers and studies, the latter to practical wisdom and storytelling. This divergence reflects the distinct audiences and purposes of each literature. For the purposes of our study, there is no contradiction in content, but rather in emphasis and form.

\paragraph{\textbf{Granularity and Conceptualization:}} Academic research tends to conceptualize and subdivide "communication" into more refined categories, discussing, for instance, "synchronous vs. asynchronous communication," "communication in the requirements domain vs. communication in code coordination," or "intercultural communicative competence." These nuances rarely appear in blogs or opinion articles, which typically treat communication more generically. Gray literature frequently lumps everything under the umbrella of "soft skills" or "communication and teamwork" without clearly delineating types.

\paragraph{\textbf{Contextualization of English and Culture:}} We note that the concern with English communication and intercultural competence appears prominently in the international academic literature (given the reality of global teams and outsourcing). Studies emphasize that fluency in technical English and the ability to engage in intercultural communication are important differentiators, especially for non-Anglophone countries working on distributed projects. The gray literature focused on communication did not place such explicit emphasis on this—possibly because many blogs already assume English proficiency (since they are written in English) or because this aspect is treated separately in discussions about ``English for programmers.'' The Portuguese content analyzed, e.g., Lancetti et al.\cite{lancetti2023ticket}, reinforces the importance of English and cultural respect, but this is an article in proceedings. Curiously, Brazilian developer blogs talk a lot about the need for English, but as a skill separate from ``communication'' (focusing on reading documentation, etc.). Thus, there is a slight divergence in emphasis: academia incorporates language and culture within the communication package, while practitioners often treat language as a separate competence.

\paragraph{\textbf{Breadth of Effects Considered:}} When discussing the consequences of (not) having communication skills, the sources also vary in scope. Academic research usually links communication to project performance, employability, and learning. The gray literature, in addition to these, adds impacts on daily dynamics and job satisfaction. For example, a widely discussed aspect in professional forums is that communicating well makes the work environment better, reduces friction, increases trust among colleagues, etc., which boosts morale and talent retention. Academic literature tends not to capture "morale" or "climate" so directly (they are difficult to measure, although some qualitative studies mention them). Therefore, professionals also emphasize the human and well-being side: ``Teams that communicate well enjoy working together more,'' which does not always appear in articles focused on productivity metrics or project success. This is not a contradiction, but a difference in focus: academia looks at tangible results (productivity, software quality), while the industry also perceives intangible results (pleasant environment, personal growth).

In general terms, we did not identify conflicting divergences in the central content—that is, there is no case where one side says "communication is crucial" and the other says "it is not that important." The divergences are more about perspective and detail. Garousi \& Rainer\cite{garousi2022introduction}, when discussing the relationship between gray and academic literature, point out that each complements the other—academia brings rigor and conceptual models, while the industry brings topicality and applicability.
Our study precisely reflects this in the context of communication: we found complementarity, not opposition.

In conclusion of RQ2, we would say that the academic and gray knowledge bases converge in the fundamental message that communication skills are indispensable in Software Engineering, diverging only in the style of presentation and contextual emphases. This finding reinforces the reliability of the knowledge: when both systematic research and practical wisdom point in the same direction, we have certainty that investing in communication is beneficial for software professionals and organizations.

\section{Discussion}

The results of this multivocal review highlight that the topic of ``communication skills in software engineering'' is a point of rare alignment between theory and practice. Communication emerges as a common denominator—universally valued and treated as a strategic competence in software development. In this section, we discuss the implications of these findings, explore possible reasons for the identified divergences, and offer critical interpretations in light of the area's current context.

\paragraph{\textbf{Why is there so much convergence?}} Unlike technical topics where academia and industry sometimes diverge (e.g., adoption of certain methodologies or tools), in the case of soft skills, both sides daily experience the human challenges of projects. It is reasonable to assume that SE researchers are also, for the most part, software professionals or ex-professionals, or they interact with companies to conduct studies. Thus, there is a more direct exchange of perceptions. Furthermore, communication problems are easily recognizable. It is not uncommon for project managers or professors to witness a project fail due to a misunderstanding. This tangibility facilitates consensus. As Garousi et al.\cite{garousi2020gray} pointed out, the lack of communication between communities harms both; curiously, communication is precisely the remedy pointed to for integrating academia and practice. Our findings suggest that, at least conceptually, this integration begins with agreement on its importance.

\paragraph{\textbf{Different Perspectives:}} The divergences in emphasis observed can be interpreted as a result of each community viewing the problem from different perspectives. Academia applies a scientific viewpoint: it needs to define, categorize, and measure to produce generalizable knowledge. The industry, on the other hand, has a more pragmatic viewpoint: it needs to solve immediate problems and improve effectiveness in the short term. Therefore, academia speaks of "interpersonal, written, intercultural communication competencies..." while the industry says "communicate well and be clear." In essence, both are correct, just at different levels of analysis.

\paragraph{\textbf{Leveraging Complementarity:}} Given the evidenced complementarity, an opportunity arises to combine insights from both sides for more robust solutions. For example, corporate training programs could leverage tested academic methodologies (evidence-based communication workshops, 360° evaluations, etc.) while maintaining accessible language and practical examples to engage participants. Likewise, academic curricula can invite industry professionals to share real-world cases of communication problems, making learning more concrete for students. This rapprochement aligns with recent recommendations for multivocality, where guidelines encourage researchers to consider blogs and other artifacts in research formulation, and practitioners to engage in collaborative research.

\paragraph{\textbf{Practical Implications:}} For professionals and project managers, our results convey a clear message: investing in communication skills is not optional, but rather critical for achieving success throughout Software Engineering. This implies paying attention to continuous training, for example, encouraging developers to give internal tech talks to practice technical communication, promoting an open feedback culture (perhaps with feedback workshops), and recognizing good communication behaviors in performance evaluations. Companies can include communication criteria in promotions and hiring more systematically. For educators and curriculum designers, the convergence between industry and academia on this point legitimizes increasing the workload of activities focused on soft skills without fear, as it is no longer about "nice-to-have" or "it will be a differentiator," but rather "necessary-to-have."

\paragraph{\textbf{Measurement is still a concern:}} A challenge highlighted in the sources (especially in the industry) is how to measure and evaluate communication. While it is easy to test programming with a code challenge, assessing whether someone communicates well is more subjective. Some organizations use group dynamics or situational questions in interviews to infer this. Academic literature suggests competence frameworks that can help clarify criteria. For example, evaluating communication in terms of: clarity, objectivity, empathy, listening. Still, it is a contextual skill—someone might communicate well in writing but not orally, and vice versa. This indicates that, although there is consensus on its importance, there is still ground to explore, not only on how to develop but also on how to evaluate this skill. This is a fruitful frontier for academia-industry collaboration: controlled experiments on communication training techniques and their impact on agile teams, for example.

\paragraph{\textbf{Impact of Local Culture:}} A point not deeply explored in the sources, but noteworthy, is how cultural differences can influence the perception of ``good communication.'' In more direct vs. indirect cultures, the form of feedback changes—something seen as clear communication in the US could sound rude in Japan, for example. International academic literature touches upon this by discussing ``intercultural competence''; the gray literature rarely enters into this merit, tending to assume Western contexts. For a global company or distributed remote teams, this nuance is vital: what constitutes good communication can vary culturally.

\subsection{Limitations and Threats to Validity}

It should be noted that our review, while comprehensive, does not cover all existing literature. It is possible that there are isolated academic studies that do not find a significant correlation between communication and success (perhaps due to study design) or niche blogs that question the emphasis on soft skills. However, we believe these would be outliers. The strong convergence observed here indicates a dominant trend. Furthermore, our own interpretation might be influenced by confirmation bias, given that we started with a preconception that communication is important, and thus we may have paid less attention to counter-examples. We sought to mitigate this by actively looking for alternative views, but we found virtually none explicitly in the sources. Also, the qualitative nature of the comparison leaves room for nuance: we emphasize that divergences are in tone and focus, but another researcher might classify them differently. We sought to base each statement on cited evidence from the sources, for transparency.

\section{Conclusion}

This paper investigated how communication skills in Software Engineering are discussed and valued across two (complementary) spheres of knowledge—academic literature and gray literature—and analyzed the convergences and divergences between them. In response to RQ1, we found that, overall, communication skills are consistently portrayed as essential in Software Engineering. Academic literature presents evidence of its importance, showing communication as one of the most demanded soft skills by employers and associated with better project outcomes. Gray literature reinforces this view with practical examples and emphatic advice, considering communication a differentiator for individual, collective, and project success, whether in internal team collaboration or in dealing with clients. Both sources highlight multiple aspects: clarity in expressing ideas, active listening, constructive feedback, comprehensible documentation, among others, all forming the communicative repertoire necessary for software professionals.

In relation to RQ2, we conclude that there is a predominance of convergences between the academic and practical perspectives, with some divergences of nuance. Both domains agree on the central premise: effective communication is valued as a pillar in software development environments and must be developed just as much as technical skills. Convergences include the identification of the same critical sub-skills (oral, written, collaboration, etc.) and the recognition that communication enhances productivity, which consequently impacts project quality. The divergences primarily reside in the tone and approach. Academia addresses the topic systematically and evidentially, while the industry does so experientially and motivationally—and in some contextual emphases, such as the issue of language/culture, which is more explicitly treated in formal sources. Nevertheless, these differences do not represent contradictions about what is important, but rather how the subject is articulated.

\section*{Availability of Artifacts}

All supplementary materials necessary for the replicability of this research are
available in the repository: \url{https://figshare.com/s/13ffc6100bb9539d22e1}

\section*{Acknowledgments}
This study was partially funded by CAPES - Finance Code
001, INES.IA (National Institute of Science and Technology for Software Engineering Based on and for Artificial Intelligence, \url{www.ines.org.br}), and CNPq (408817/2024-0, 314680/2026-8 and 308623/2022-3).

\bibliographystyle{unsrt}
\bibliography{sbc-template}

\end{document}